# Visualizing Graphene Based Sheets by Fluorescence Quenching Microscopy

*Jaemyung Kim, Laura J. Cote, Franklin Kim, and Jiaxing Huang\**

Department of Materials Science and Engineering, Northwestern University, Evanston, Illinois 60208, USA

\* Email: Jiaxing-huang@northwestern.edu

**ABSTRACT**

Graphene based sheets have stimulated great interest due to their superior mechanical, electrical and thermal properties. A general visualization method that allows quick observation of these single atomic layers would be highly desirable as it can greatly facilitate sample evaluation and manipulation, and provide immediate feedback to improve synthesis and processing strategies. Here we report that graphene based sheets can be made highly visible under a fluorescence microscope by quenching the emission from a dye coating, which can be conveniently removed afterwards by rinsing without disrupting the sheets. Current imaging techniques for graphene based sheets rely on the use of special substrates. In contrast, the fluorescence quenching mechanism is no longer limited by the types of substrates. Graphene, reduced graphene oxide, or even graphene oxide sheets deposited on arbitrary substrates can now be readily visualized by eye with good contrast for layer counting. Direct observation of suspended sheets in solution was also demonstrated. The fluorescence quenching microscopy offers unprecedented imaging flexibility and could become a general tool for characterizing graphene based materials.

**Introduction**

Microscopy imaging techniques usually play a critical role in materials discoveries at small length scales. For example, the discovery that graphene is visible under a normal optical microscope when deposited on dielectric-coated silicon wafers[1,2] has enabled numerous studies on these single atomic carbon sheets[3-5]. Graphene oxide (G-O)[7,8] is a promising precursor for solution processed, chemically modified graphene (*a.k.a.* reduced G-O) thin films for applications such as flexible, transparent conductors[9-11]. Since the optical absorption of G-O is much weaker than graphene[12], it is even more challenging to visualize under optical microscope. Atomic force microscopy (AFM) is often used to visualize G-O sheets since it gives accurate thickness value at nanometer scale[13]. However, it has not been made suitable for quick sample examination over large areas due to rather low throughput. In addition, it typically requires very smooth substrates such as freshly cleaved mica or silicon wafer. Scanning electron microscopy (SEM) can be much faster but it needs to operate under vacuum and works best for films deposited on conducting substrates[14]. High-contrast optical imaging of G-O sheets has also been demonstrated by interference technique[15] and imaging ellipsometry[16], but only on dielectric-coated silicon wafers, where the thickness of the dielectrics and the illuminating wavelength need to be optimized. The need for special types of substrates to visualize graphene based sheets greatly limits our capability to study these new two-dimensional sheets. For example, solution processed graphene films are found to be promising for flexible, transparent plastic electronics. To establish how processing conditions affect on the final thin film quality, it is critical to see the microstructures of the film to find out the size distribution of the sheets, the coverage on the plastic substrate, and the degrees of wrinkling and overlapping. However, imaging graphene based sheets deposited on plastic surface has been a great challenge with current microscopy techniques. Therefore, alternative methods that can image graphene based sheets without the need for special substrates would be very useful for high-throughput sample evaluation in diverse applications. Here we report a general method for visualizing graphene based sheets on arbitrary substrates by fluorescence quenching microscopy (FQM). The fluorescence quenching mechanism eliminates the need for special substrates and even allows the direct observation of graphene based sheets in solution. It offers unprecedented imaging flexibility for characterizing graphene based materials.

A recent publication by Treossi *et al.*[17] has shown that G-O sheets can be visualized on glass, quartz, and silicon through quenching the fluorescence of a thiophene dye covalently tethered to the substrates. The current work presents an enhancement because our FQM method can produce layer contrast, does not involve surface functionalization thus allowing observation on arbitrary substrates, and enables real-time solution phase imaging.

**Experimental Section**

**Synthesis of graphene, G-O, and reduced graphene oxide (r-G-O).** Graphene was prepared by micro-mechanical cleavage of highly oriented pyrolytic graphite using "Scotch Tape" method[1]. G-O was synthesized using a modified Hummers and Offeman's method from graphite powder (Bay carbon, SP-1)[14,18,19]. Chemically reduced graphene oxide (r-G-O) was prepared by exposing G-O coated substrates to hot hydrazine vapor (Sigma Aldrich, anhydrous, 98%) in a sealed chamber maintained at 80°C for overnight. Although all these types of sheets were successfully visualized, G-O sheets were used in most experiments because they are a much weaker absorber and less effective quencher than r-G-O or graphene, and therefore represents a "worst case" scenario for FQM imaging

**Solution phase fluorescence quenching measurement.** r-G-O water dispersion was prepared by hydrazine reduction of G-O[10]. Fluorescence spectra of fluorescent dye solutions were acquired before and after adding minute aliquots of G-O or r-G-O dispersions. The volume and concentration of G-O and r-G-O dispersions added were kept the same. 3 dyes with very different molecular structures and absorption/emission profiles were tested including a red florescent dye 4-(dicyanomethylene)-2-methyl-6-(4-dimethylaminostyryl)-4H-pyran (DCM, Sigma-Aldrich, 98%), a green fluorescent dye fluorescein sodium salt (Sigma-Aldrich), and a blue fluorescent dye 2,5-bis(5-tert-butyl-2-benzoxazolyl)thiophene (BBOT, TCI America, >98%). The fluorescence spectra were obtained by a photon counting spectrofluorimeter (ISS, PC1).

**Sample preparation.** Glass microscope coverslips (VWR) and $SiO_2$/Si wafers were cleaned following standard RCA treatment method. Polyester substrates (Eppendorf) were cleaned with deionized water. G-O film was deposited by Langmuir-Blodgett technique[14] (Nima Technology, Medium size LB deposition trough), spin-coating (Laurell Technologies Corporation, WS-400, 1 min at 4000 rpm), or drop casting. To improve the uniformity of the dye coating, a polymer was co-dissolved with the dye for spin coating. Typically, 1 mg of a green fluorescent dye - fluorescein sodium salt powder was added to 10 ml of polyvinylpyrrolidone (PVP, Sigma-Aldrich, $M_W$ = 55,000)/ethanol solutions. Solutions with 0.1, 0.5, 1, and 5 wt% of PVP were prepared to vary the thickness of the coating. Since PVP forms a charge transfer complex with fluorescein sodium salt at high polymer concentration,[20] for the 5 wt% PVP solution, 2 mg of dye powder was added to compensate the fluorescence quenching by PVP. 100 μl of the coating solution was dispensed for each 0.5 $in^2$ of substrate area, and spun for 5 sec at 300 rpm and then 45 sec at 4000 rpm. The films produced from 0.5, 1, and 5 wt% of PVP solutions were measured to be approximately 10 nm, 30 nm, and 200 nm thick by surface profilometer (Veeco, Dektak 150), respectively. The thicknesses of films produced from 0.1 wt% PVP solution were found to be smaller than 5 nm, although the exact values were difficult to determine due to intrinsic surface roughness of the coverslips. The dye/polymer film was also prepared with resist materials that are commonly used in photolithography and e-beam lithography for device fabrication such as SU-8 (Microchem, 2000.5) and poly(methyl methacrylate) (PMMA, Sigma-Aldrich, $M_W$ = 120,000). 0.01 wt% of DCM was added to 10 ml of 0.5 wt% PMMA/chloroform solution. Then the solution was dispensed upon a substrate drop-wise (100 μl for each 0.5 in$^2$ substrate area) while spinning at 8000 rpm for 1 min. For SU-8, it was first diluted with ethyl L-lactate (Alfa Aesar, 99%) to a volume ratio of 1:4 (SU-8 : ethyl L-lactate), and then mixed with the same volume of 0.02 wt% DCM/ethyl L-lactate solution. Spin coating was done at 3000 rpm for 1 min with 100 μl of the solution for each 0.5 in$^2$ substrate area. The thickness of both PMMA and SU-8 coating were measured to be approximately 25 nm by surface profilometer.

**Fluorescence quenching microscopy (FQM).** FQM was performed on a Nikon TE2000-U inverted fluorescence microscope with the Exfo X-cite illumination system using an ET-GFP filter cube (FITC/Cy2, Chroma Technology Corp) for most of the experiments. Most images were taken by a monochrome interline CCD camera (Photometrics, CoolSNAP HQ$^2$) unless otherwise mentioned. The image contrast was defined as $C = (I_B - I_G)/I_B$, where $I_B$ and $I_G$ are the brightness of the background and the graphene based sheets in a FQM image, respectively. Values of brightness were read from 10 randomly chosen spots from G-O single layers and another 10 spots from background, and then averaged to calculate $C$.

To test the remote fluorescence quenching hypothesis, a non-fluorescent polymer layer was applied to separate the G-O sheets and the dye layer. Polystyrene was chosen as the spacer layer since it can form an immiscible bilayer with PMMA by spin coating[21]. In these experiments, 0.5 and 5 wt% of polystyrene (PS, Scientific Polymer Product, $M_W$ = 45,000)/toluene solutions were prepared and spin-coated onto a RCA treated glass coverslip at 3000 rpm for 1 min (100 μl for each 0.5 in$^2$ substrate area). The thickness of the film was measured to be approximately 20 nm and 200 nm, respectively, by profilometry. Then a DCM doped PMMA layer was spin-coated on top of the PS film from 0.02 wt% DCM/0.1 wt% PMMA solution in acetic acid (EMD chemicals, glacial ACS) to create a fluorescent coating that was measured to be a few nanometers thick. A sample without PS underlayer was also prepared as a control. The illuminating intensity and camera exposure time were maintained constant for FQM imaging of each sample.

Solution phase observation was conducted with a droplet of G-O/fluorescein solution confined between two glass coverslips. Both water and methanol were used as solvent. Similar dewetting behaviors were observed for both solvents. Images in Figure 9a and 9b were taken with a water droplet. Snapshots in Figure 9c and 9d were taken with a methanol droplet, which evaporates faster so that extended period of time can be avoided for recording a complete dewetting event. To avoid excessive photo-bleaching, lowest illumination intensity of the light source (12%) was used. Better resolution was observed with higher level of illumination. All FQM images presented in the paper, except the snapshots in Figure 9d, were as-acquired without further adjustment in contrast or brightness.

**Characterization by other microscopy technique.** AFM images were acquired on a scanning probe microscope (Veeco, MultiMode V). Bright field optical images were taken with a CCD camera (Diagnostic Instruments Inc., SPOT Insight QE 4.2) on a Nikon E600 upright microscope.

## Results and Discussion

It has been well known that graphitic systems such as carbon nanotubes[22,23] can strongly quench the emission of nearby dye molecules through energy transfer. Graphite itself has been used to reduce fluorescence interference in Raman spectroscopy[24]. Recent theoretical[25,26] and experimental[27] studies showed that graphene should also be a highly efficient quencher. Indeed, the fluorescence spectra of three different dyes with distinct molecular structures and absorption/emission profiles (Figure 1) showed that the emission of dye solution can be significantly quenched by adding a small aliquot of r-G-O (black line) or even G-O (brown line), suggesting that the quenching effect is general to fluorescent materials. The strong quenching by G-O is likely due to the residual graphitic domains in the basal plane that survived the severe chemical oxidation[28-30]. Among the three dyes, fluorescein was chosen as a model dye compound due to its low cost and high quantum yield.

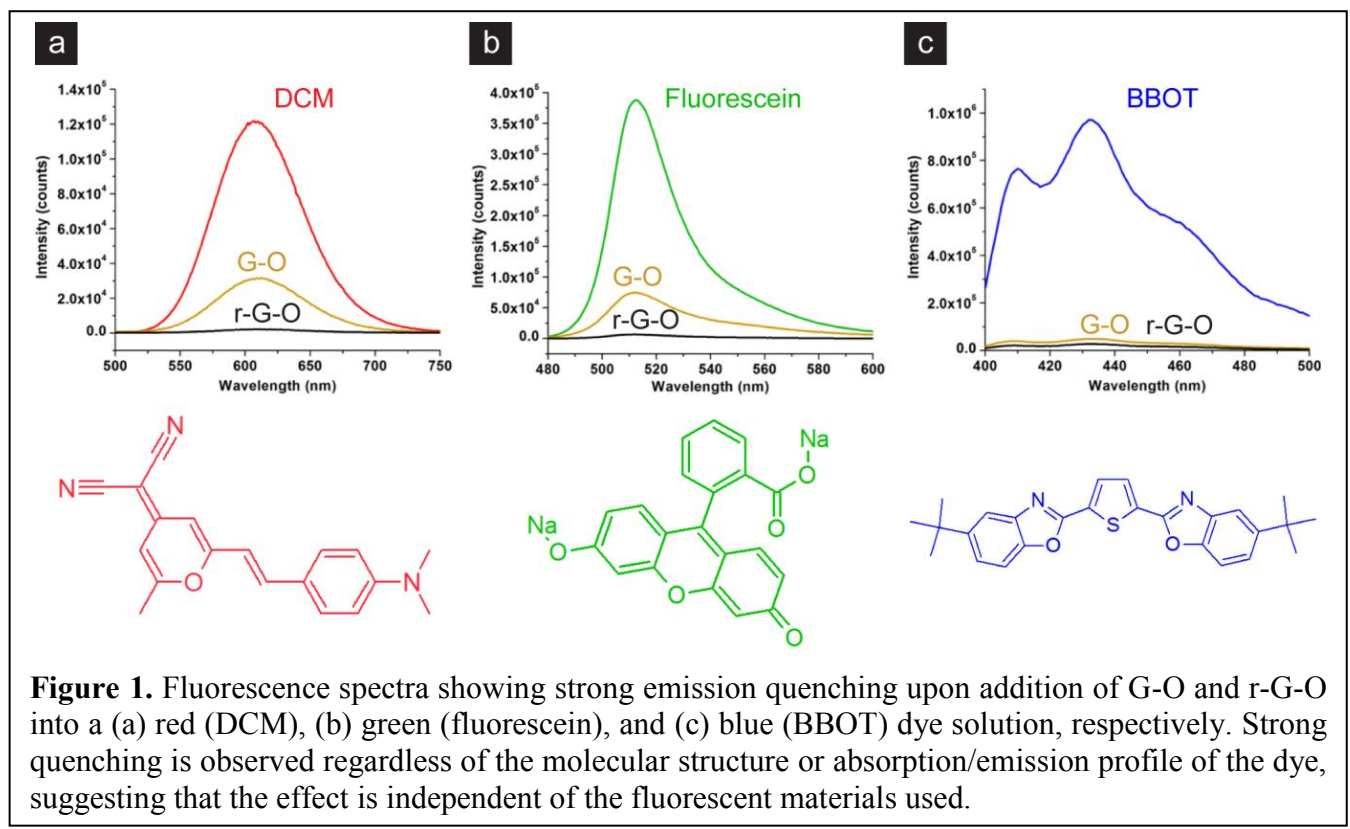

**Figure 1.** Fluorescence spectra showing strong emission quenching upon addition of G-O and r-G-O into a (a) red (DCM), (b) green (fluorescein), and (c) blue (BBOT) dye solution, respectively. Strong quenching is observed regardless of the molecular structure or absorption/emission profile of the dye, suggesting that the effect is independent of the fluorescent materials used.

Figure 1 inspired us to develop fluorescence quenching microscopy (FQM), utilizing emission quenching as a contrast mechanism for visualizing graphene based materials (Figure 2a). Typically this can be achieved by spin-coating with a fluorescein/ethanol solution. A soluble polymer, such as polyvinylpyrrolidone (PVP) was added to the solution to improve the uniformity of the resulting film. A test sample was prepared, in which both G-O and r-G-O sheets were deposited on the same glass coverslip. First, a G-O film was deposited on half of the coverslip by Langmuir-Blodgett (LB) technique, and reduced to r-G-O by hot hydrazine vapor. Then the substrate was rotated by 90° to collect a second G-O layer. The crossed depositions thus created four quadrants on the coverslips that can be easily identified as G-O, G-O/r-G-O, r-G-O and blank domains. The picture in Figure 2b shows that the emission from a fluorescein solution (Figure 2b, left) was significantly quenched upon the addition of small amount of r-G-O (Figure 2b, right), or even G-O (Figure 2b, middle). Figure 2c is the FQM image of the test sample after applying a fluorescein/PVP coating. The high image contrast allows easy identification of G-O and r-G-O sheets and clearly reveals the four quadrants of the coverslip. The r-G-O sheets indeed appeared darker than the G-O sheets, which is consistent with their higher quenching efficiency (Figure 1b and 2b). Since G-O is a less effective quencher than r-G-O or graphene, we deliberately chose it as the model material in the subsequent experiments as it should represent the "worst case" scenario for FQM imaging.

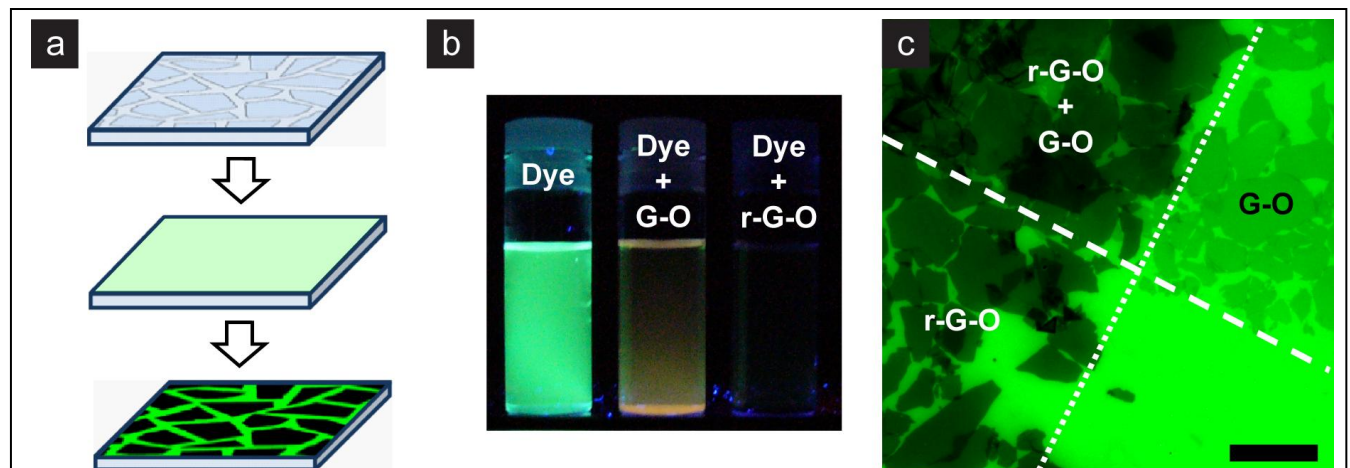

**Figure 2.** Visualizing graphene based single atomic layers by fluorescence quenching microscopy (FQM). (a) In FQM, a dye coating is applied to a graphene, G-O or r-G-O covered surface, which upon excitation reveals the underlying sheets due to fluorescence quenching. (b) A camera image showing strong emission quenching upon addition of G-O and r-G-O to a fluorescein solution. (c) A cross-deposited G-O/r-G-O sample on glass coverslip showing four quadrants of G-O, r-G-O/G-O, r-G-O and blank domains (counterclockwise). r-G-O sheets appeared darker due to higher quenching efficiency. Scale bar = 25 μm.

In order to verify that FQM can indeed visualize single layers, we compared FQM and AFM images of the same G-O sheets. Figure 3a is an AFM image showing a few G-O sheets deposited on a SiO$_2$/Si wafer. Height measurements (Figure 3d) confirmed that they were single layers of around 1 nm in thickness[13]. The height of folded areas was measured to be around 2 nm. Figure 3b is an as-acquired FQM image of the same area after applying a 30 nm thick fluorescein/PVP layer. It perfectly matches the AFM view in Figure 3a with clear contrast between single and double layers, suggesting higher degree of quenching by multilayers. Like SEM, FQM does not offer absolute measurement of the number of layers. For G-O sheets, the layers with the smallest contrast were assumed to be single layers. One may ask whether FQM would tell if a G-O sample does not have single layers, but only multilayers. If the multilayers are composed of perfectly overlapped sheets with identical shape and size within optical resolution, under the above mentioned assumption FQM would mistaken them as single layers. However, while this scenario might be encountered with mechanically exfoliated, or CVD synthesized graphene samples, it is highly unlikely for G-O due to their irregular sizes and shapes, and strong electrostatic repulsion between sheets, which typically lead to partially overlapped and wrinkled multilayers.[14] Although a dye coating is needed for FQM imaging, it can be easily removed by brief washing with ethanol or water afterwards without disrupting the underlying sheets. The AFM image of the same G-O sheets after dye removal (Figure 3c) appears identical to the one before applying the dye layer. So does the height profile of the folded area (Figure 3d). No contamination or change in sheet morphology can be detected in both Figure 3c and 3d. The high contrast of FQM allows comfortable naked-eye observation without the need for special cameras. An image taken with a cheap consumer digital camera through the eye piece is shown in Figure 4, in which vivid details of the sheet morphologies can be seen.

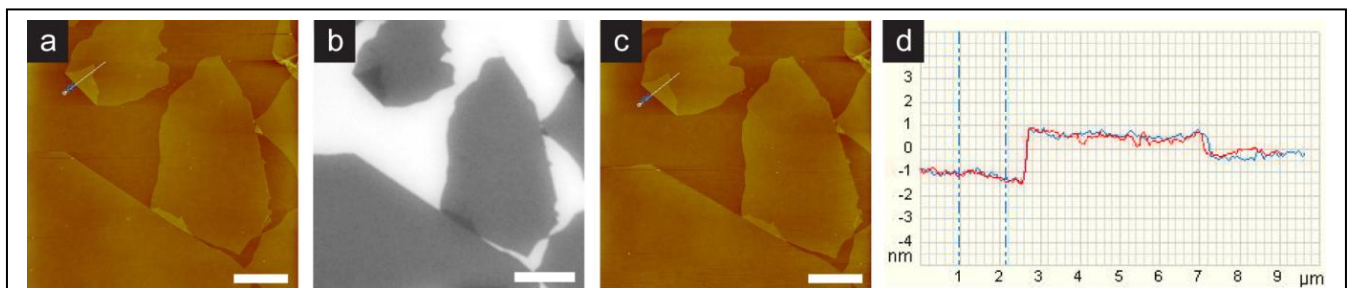

**Figure 3.** (a) AFM image showing G-O single layers deposited on a SiO$_2$/Si wafer before applying a 30 nm thick fluorescein/PVP layer for FQM. (b) A FQM image of the same area of the wafer, showing good correlation to the AFM view. (c) After washing off the dye coating, no residues can be detected by AFM. (d) Line scan data on a folded sheet show no significant deviation in thickness before and after FQM imaging. All scale bars = 10 μm.

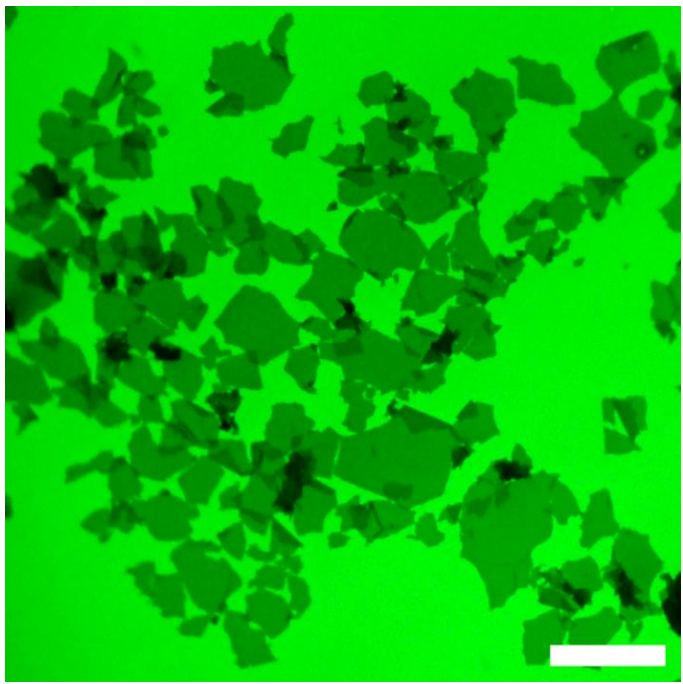

**Figure 4.** A FQM image of G-O sheets on a glass substrate with a 30 nm thick fluorescein/PVP coating taken by a cheap consumer digital camera (Panasonic, DMC-FZ1). This image is representative to what can be seen by naked eye with FQM. Scale bar = 50 μm. The green background was from fluorescein emission.

FQM can also visualize mechanically exfoliated graphene sheets as shown in Figure 5c. The single and multi-layer domains deposited on SiO$_2$/Si substrate were clearly resolved, correlating well with images taken by AFM (Figure 5a) and the commonly used bright field, reflective optical microscopy (Figure 5b). However, we noted that it was much easier to find graphene sheets using FQM due to higher contrast. FQM is not limited by the wavelength of illumination. Many fluorescent materials are available in case a specific excitation wavelength is preferred. In addition, a great variety of film forming polymers, even resist can be used as the coating layer. Figure 6a and 6b show FQM images of G-O monolayer on glass coverslip, coated with an approximately 25 nm thick film of DCM doped SU-8 and PMMA, respectively. SU-8 and PMMA are widely used resist materials for photolithography and e-beam lithography during device fabrication[31,32]. DCM can be excited by green light, which is safe for the SU-8 or PMMA resist. It is worth noting that device fabrication on graphene based sheets already relies on the use of resist materials during lithography steps. Therefore, the success of dye doped resist materials as the fluorescent layer makes our FQM technique well compatible with current microfabrication techniques. This suggests that the scope of "on-sheet" microfabrication of graphene based devices, which has usually been done on dielectric-coated silicon wafers, can be significantly broadened by FQM.

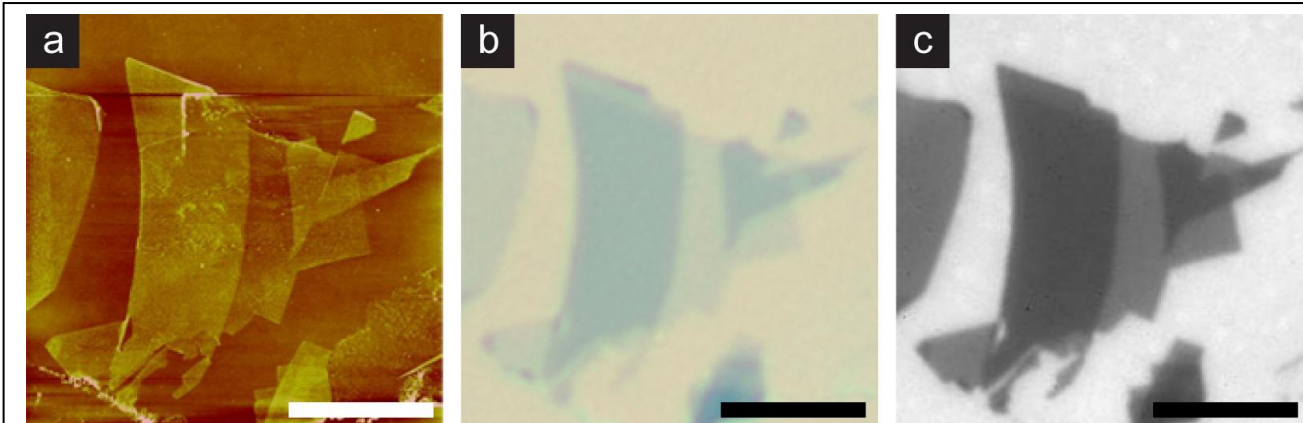

**Figure 5.** Images of mechanically exfoliated graphene on a SiO$_2$/Si substrate taken by (a) AFM, (b) optical microscopy, and (c) FQM using PVP/fluorescein. All scale bars = 10 μm.

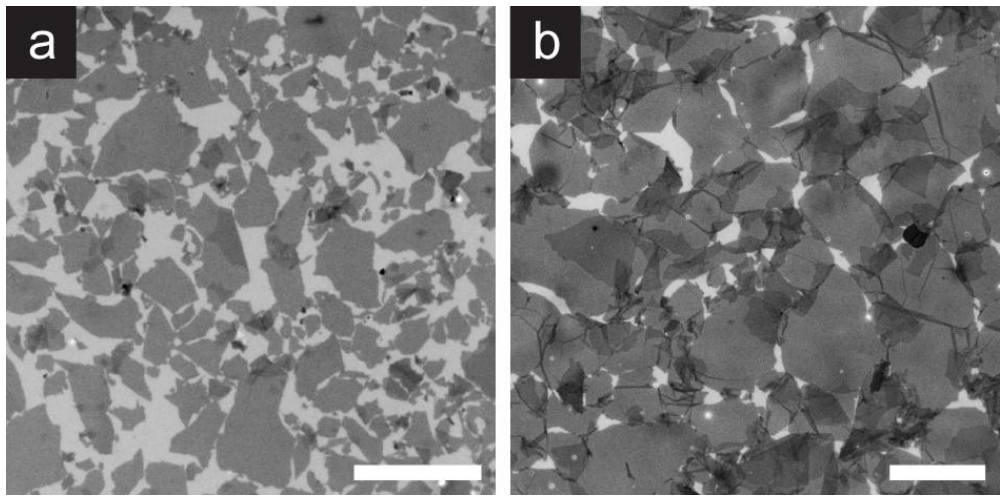

**Figure 6.** FQM images obtained with a 25 nm thick layer of (a) SU-8, and (b) PMMA, respectively, doped with red fluorescent DCM dye. Scale bars = 50 μm.

The contrast of FQM originates from fluorescence quenching by graphene based sheets, which creates dark regions in the bright dye/polymer layer upon excitation. The visibility contrast can be described as $C = (I_B - I_G)/I_B$, where $I_B$ and $I_G$ are the optical intensities of the background and the graphene based sheets in a FQM image, respectively. $C$ is essentially a measure of percentage quenching (Figure 7). For dye layers thinner than 5 nm, nearly 100% quenching ($I_G \approx 0$) was observed, leading to nearly full contrast of $C=0.98$ (Figure 7c). However, single and multilayers could not be distinguished due to "oversaturated" contrast. This suggests that the underlying G-O single layer can effectively quench the emission of nearly all the dye molecules above it in the polymer film. A recent theoretical study on remote quenching of dye molecules near graphene surface suggested that the effective quenching distance could extend to around 30 nm through resonance energy transfer[26]. Although G-O is a much weaker quencher, it appears that it is capable of quenching the emission of dye molecules that are several nanometers away. This indicates that the effective quenching distance of fluorescein near G-O surface in the PVP matrix is likely to be greater than 5 nm (Figure 7a). With a thicker coating (Figure 7b), the outer most part of the dye materials could become beyond the "reach" of the G-O sheets, which will remain bright upon excitation. This should decrease the overall contrast since $(I_B - I_G)$ is determined by the effective quenching distance, which should remain nearly constant, while $I_B$ increases with thickness. Indeed, this trend was confirmed in Figure 7c-f. When the thickness of the dye layer was increased to 10, 30 and 200 nm, the single layer contrast decreased to 0.68, 0.53 and 0.24, respectively. However, the difference between single and multilayers now became more apparent, making it possible to do layer counting. The optimal thickness that allows comfortable naked-eye observation was found to be in the range of 20 to 50 nm. The concentration of dye molecules in the polymer film determines the brightness of the background in FQM images. When dye concentration was too low (<0.0025 wt%), the images were too dim for naked eye observation. Self-quenching of dye molecules[33] was observed with dye concentration higher than 2.5 wt%. The optimal range of fluorescein concentration in the spin coating solution was found to be around 0.01 ~ 0.02 wt%. Similar effects of layer thickness on FQM visibility were observed with graphene and r-G-O samples.

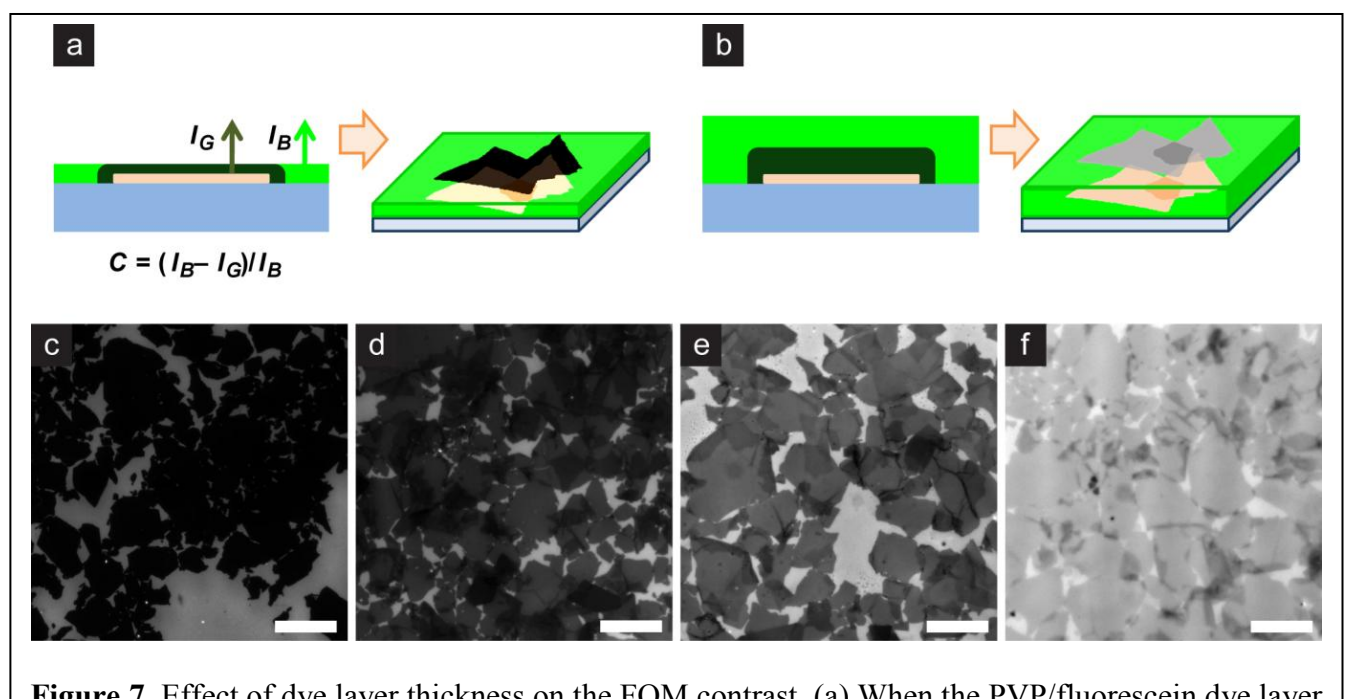

**Figure 7.** Effect of dye layer thickness on the FQM contrast. (a) When the PVP/fluorescein dye layer (green) is thinner than the effective quenching distance (black), emission from the entire dye layer can be quenched by G-O (pink yellow), leading to near full contrast *C*≈1. This would result in over-saturated contrast making multilayers indistinguishable from monolayer. (b) With thicker dye layer, FQM image contrast should decrease due to higher background fluorescence. However, this is beneficial for layer-counting as multilayers should appear darker. From (c) to (f), the thickness of dye layer was altered from less than 5 nm, to 10 nm, 30 nm and 200 nm, respectively, by changing the concentration of PVP in the coating solution. (c) With the thinnest coating (<5 nm), both the sheets and their overlapped regions appear black. (d-f) As the thickness of dye layer was increased, the image contrast decreased but single and multilayer domains became distinguishable. The optimal thickness of dye coating was found to be around 20 to 50 nm. All the G-O films were deposited on glass slides. All scale bars = 50 μm.

The effect of dye layer thickness on FQM contrast suggests that G-O sheets can indeed quench the emission of dye molecules remote to its surface. Therefore, if a non-fluorescent spacer layer is placed between the G-O sheets and the dye coating, the contrast of FQM should decrease. To test this hypothesis, we employed a bilayer coating of PS/PMMA on G-O where PS was the spacer, and DCM doped PMMA was the fluorescent layer. This was done by consecutive spin coating steps of each polymer solution on a G-O coated glass substrate following known procedure for creating immiscible PS/PMMA bilayers[21]. Figure 8 shows the effect of the PS spacer on FQM contrast. Without the PS spacer (Figure 8a), the FQM image shows over-saturated contrast ($C \approx 1$) with a few nanometers thick DCM/PMMA layer. When a 20 nm PS spacer was introduced, however, the percent quenching represented by $C$ decreased to 0.23 (Figure 8b). With a 200 nm PS spacer, G-O sheets became invisible (Figure 8c). The experiments in Figure 8 strongly support the remote quenching effect of G-O, which should also be applicable for r-G-O and graphene sheets. The lower overall brightness of Figure 8a is likely due to fluorescence quenching by the substrate itself.[34]

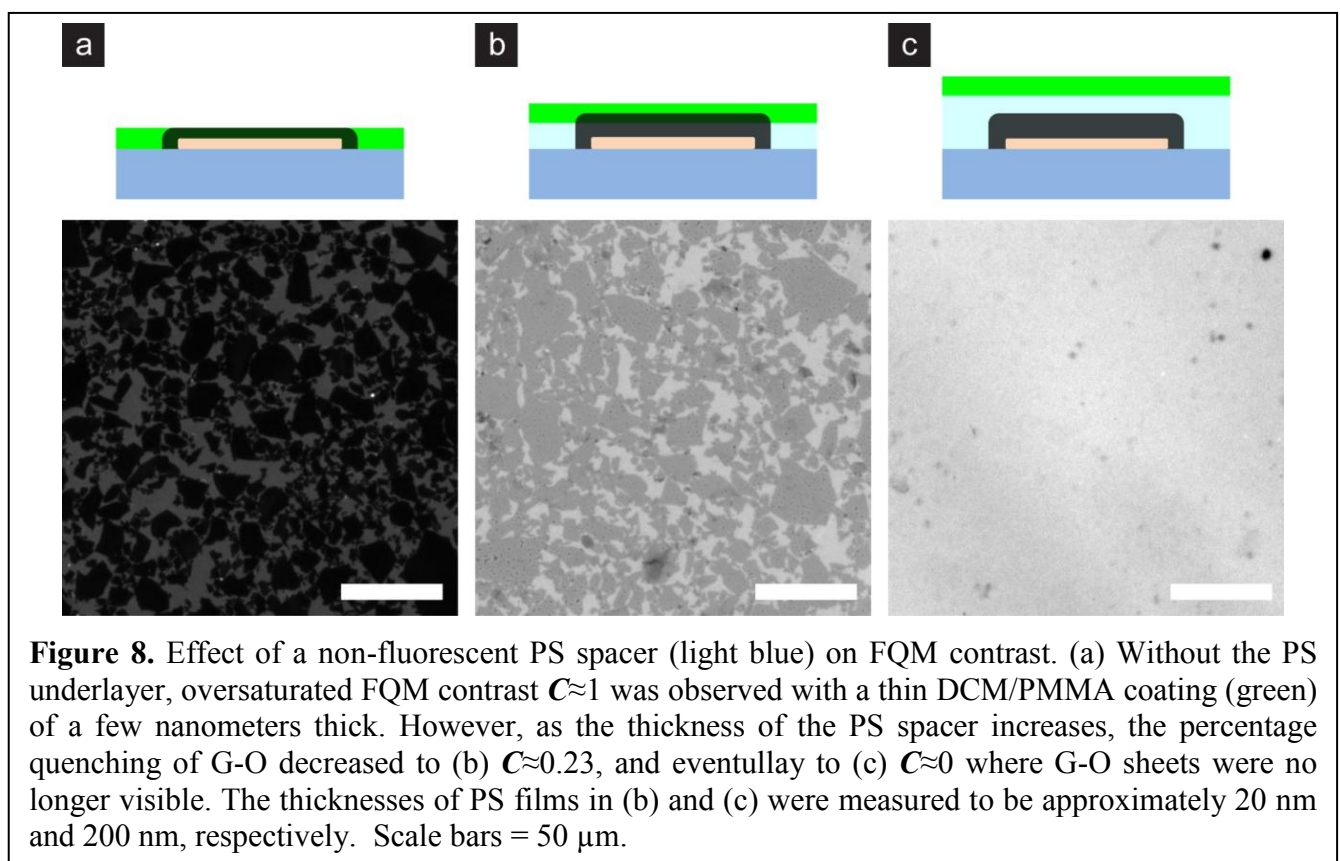

**Figure 8.** Effect of a non-fluorescent PS spacer (light blue) on FQM contrast. (a) Without the PS underlayer, oversaturated FQM contrast $C \approx 1$ was observed with a thin DCM/PMMA coating (green) of a few nanometers thick. However, as the thickness of the PS spacer increases, the percentage quenching of G-O decreased to (b) $C \approx 0.23$, and eventullay to (c) $C \approx 0$ where G-O sheets were no longer visible. The thicknesses of PS films in (b) and (c) were measured to be approximately 20 nm and 200 nm, respectively. Scale bars = 50 μm.

FQM offers unprecedented flexibility for imaging graphene based materials regardless of substrate type. Due to its simplicity, FQM can be used as a quick sample evaluation method for graphene based thin films on arbitrary substrates. For example, microscopy imaging of graphene based sheets on plastic surface has been very challenging. We have found that commonly used plastic substrates such as polyethylene microscope slides are usually too rough for acquiring good AFM images of graphene based sheets. The insulating nature of plastics also made it very difficult for SEM observation. Furthermore, direct optical imaging is hard to achieve due to the lack of a well defined dielectrics interface such as $SiO_2/Si$ or $Si_3N_4/Si$. However, FQM lifts the need for special substrates and can easily visualize graphene based sheets on plastic surface. Therefore, it can be used to evaluate how solution processing methods affect thin film morphology. Figure 9 shows G-O films deposited on polyester substrates by three different solution processing technique, namely drop casting (Figure 9a), spin coating (Figure 9b), and LB assembly (Figure 9c)[14,35], respectively. FQM was able to reveal vivid details of the wrinkles, folds, and overlaps of the sheets. On such substrates, G-O sheets could not be observed at all with a bright field optical microscope under either reflectance (Figure 9b, inset) or transmission mode. It was observed that G-O sheets deposited either by drop casting or spin coating appeared to be heavily wrinkled and folded, mainly due to uncontrolled dewetting process on polyester surface. In case of spin coating, G-O sheets were also stretched along the spreading direction of solvent. The high degree of wrinkling and folding reduces the surface coverage of G-O, therefore, increases the amount of material needed to form a continuous film. This would be an undesirable feature in transparent conductor thin film applications. In contrast, LB films had much improved surface coverage (Figure 9c).

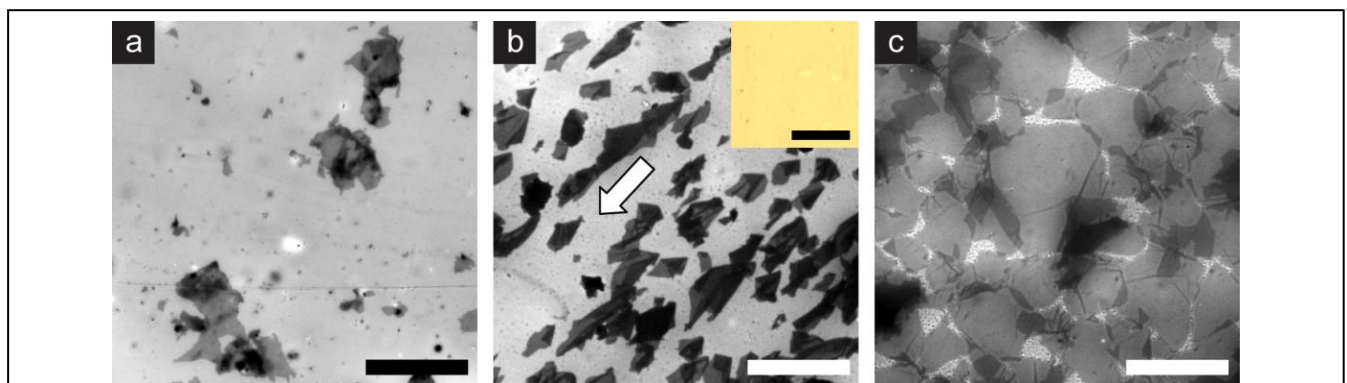

**Figure 9.** FQM evaluation of G-O sheets deposited on polyester substrates by various solution processing techniques: (a) Drop casting; (b) Spin coating and (c) LB assembly. Films shown in (a) and (b) are heavily wrinkled and folded, reducing the surface coverage of G-O. The vivid details of wrinkled sheets in (b) indicate that they were stretched along the solvent spreading direction (block arrow) during spin coating. These sheets are not visible under bright-field in reflectance mode (b, inset). (c) LB assembly produced a close-packed G-O monolayer with maximal coverage. Scale bars = 50 μm (inset = 100 μm). PVP/fluorescein was used as the coating layer.

No current technique is capable to image graphene based sheets in solution. With FQM, it is now possible to directly observe them in solution. Although G-O has weak emission in the near-infrared region, and G-O nanosheets have been used as fluorescence label for cell imaging[36], we found that the fluorescence intensity of micron-sized G-O sheets in the visible spectra was insufficient for real-time observation. However, G-O becomes highly visible as dark sheets in a dye solution upon excitation due to emission quenching. Figure 10a is a bright-field image of an evaporating droplet of G-O/fluorescein aqueous solution, in which G-O sheets are barely visible. When switched to fluorescence mode (Figure 10b), G-O sheets were revealed. The real-time imaging capability thus enables the study on their dynamic solution behaviors. As a proof-of-concept, we observed how the G-O sheets were deposited during the dewetting process since this is crucial to the final film morphology. Two typical types of behaviors of G-O at the contact line were observed, namely drifting and pinning. Figure 10c is a sequence of snapshots showing the drifting event of a G-O sheet at the meniscus. Due to its spear-like shape, the sheet rotated and wobbled until one of its longer edges was aligned with the contact line. Drifting sheets like this were usually concentrated at the center of the droplet and stacked with each other at the final stage of evaporation. Figure 10d captures the depositing process of a larger, more flexible sheet, in which it became a pinning site for the receding contact line[37]. The capillary force imposed by the meniscus thus folded the sheet into a crumpled particle after drying. These observations explain why micron-sized G-O sheets are often seen wrinkled, folded and overlapped with each other in drop-casted films, as shown in Figure 9a. FQM should also allow direct observation of many other interesting phenomena such as solvent induced conformation change of G-O[38].

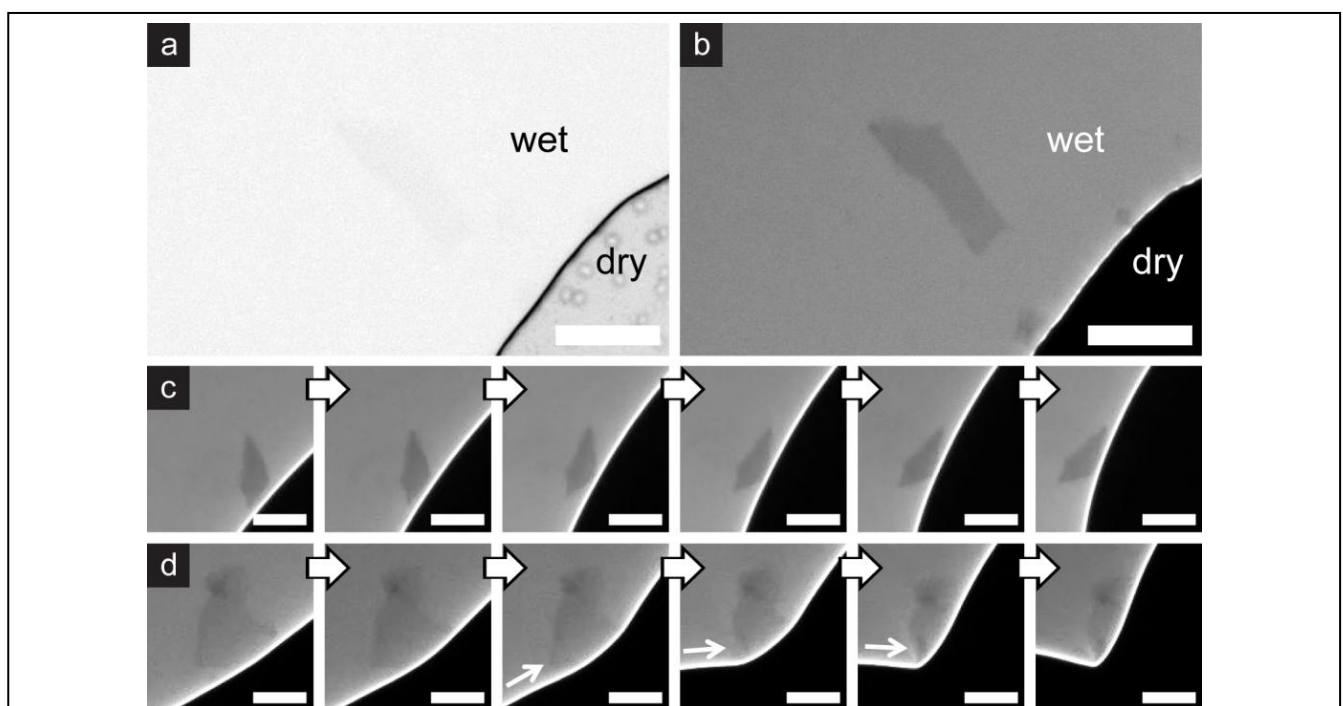

**Figure 10.** FQM observation of G-O sheets dispersed in dye solution. In contrast to (a) bright field image, G-O sheets suspended in fluorescein solution are much more visible under (b) FQM, allowing *in-situ*, real-time observation of their dynamic solution behaviors such as those during solvent evaporation. (c) In the snapshots, a spear-shaped G-O sheet was captured drifting with the dewetting front. The snapshots in (d) show contact line pinning by a depositing G-O sheet. In (a) and (b), scale bars = 30 μm. In (c) and (d), scale bars = 15 μm.

**Conclusion**

In conclusion, utilizing the strong fluorescence quenching effect, graphene based, single atomic layer carbon sheets can be visualized with a common fluorescence microscope by applying a dye doped polymer coating. The dye layer can be easily removed by washing after imaging without disrupting the underlying sheets. FQM works with a wide range of fluorescent materials and polymers including resist materials used in photolithography and e-beam lithography. This makes FQM compatible with microfabrication processes. Therefore FQM could greatly broaden the scope of single layer device fabrication since it can image these 2D sheets on arbitrary substrates. FQM enables high throughput, high contrast evaluation of graphene based sheets on plastic substrates and even in solution. The highly versatile nature of FQM should make it a general imaging tool for characterizing graphene based materials.

**Acknowledgement**

This work was supported by a Northwestern University new faculty startup fund and a seed grant from the Northwestern Nanoscale Science and Engineering Center (NSF EEC 0647560). We thank S.Y. Kim for help with initial fluorescence spectra measurement. J.K thanks the Initiative for Sustainability and Energy at Northwestern (ISEN) for graduate fellowship. L.J.C. gratefully acknowledges the National Science Foundation for a graduate research fellowship. We thank the NUANCE Center at Northwestern, which is supported by NU-NSEC, NU-MRSEC, Keck Foundation, the State of Illinois, and Northwestern University, for use of their facilities.